\documentclass{iopconfser}
\usepackage[english]{babel}
\usepackage[utf8]{inputenc}
\usepackage[T1]{fontenc}
\usepackage{amsmath,amssymb,lmodern,amsthm,accents}
\usepackage{marginnote}
\usepackage{mathtools}
\usepackage{mathrsfs}
\usepackage{bm}
\usepackage{xcolor}
\usepackage{tensor}
\usepackage{hyperref}
\usepackage{cleveref}
\usepackage{array}
\usepackage[numbers,sort&compress]{natbib}
\hypersetup{
	colorlinks=true,
	urlcolor=blue,
	linkcolor={black},
	citecolor={green!40!black}
}
\newcommand{\scri}{\mathscr{J}}
\newcommand{\ct}[2]{\tensor{{#1}}{#2}}
\newcommand{\cth}[2]{\tensor{\hat{#1}}{#2}}
\newcommand{\prn}[1]{\left(#1\right)}

\newcommand{\defeq}{\vcentcolon=}
\newcommand{\D}{\mathcal{D}}
\newcommand{\W}{\mathcal{W}}

\renewcommand{\P}{\mathcal{P}}
\newcommand{\ctr}[3]{\tensor[{#1}]{#2}{#3}}
\newtheorem*{crit*}{Criterion}

\newtheorem{thm}{Theorem}[]
\newtheorem{remark}{Remark}[]

\usepackage{color}

\newcolumntype{M}[1]{>{\centering\arraybackslash}m{#1}}
\newcolumntype{N}{@{}m{0pt}@{}}
\begin{document}

\title{Characterization of gravitational radiation at infinity with a cosmological constant}

\author{Francisco Fernández-Álvarez$^{1}$ and  José M. M. Senovilla$^{2}$}

\affil{Departamento de Física, Universidad del País Vasco UPV/EHU, Bilbao, Spain}

\email{$^1$francisco.fernandez@ehu.eus, $^2$josemm.senovilla@ehu.eus}

\begin{abstract}
The existence or absence of gravitational radiation escaping from the spacetime at $\scri$ is
characterized in the presence of a cosmological constant $\Lambda$ of any sign. To that end, the properties of
the asymptotic super-momentum are used. When $\Lambda=0$, the characterization is equivalent to that
based on the News tensor. For $\Lambda\neq 0$, it provides the first reliable definition of existence of radiation at
$\scri$, and it gives fully satisfactory results in known exact solutions.
\end{abstract}

\section{Introduction}

Gravitational radiation in full General Relativity in the absence of a cosmological constant $\Lambda=0$ was tackled decades ago \cite{Trautman58,Pirani57,Bel1962,Bondi1962,Sachs1962,Penrose62}. In more 		recent years, however, many works have dealt with the question of characterising gravitational radiation at the conformal boundary $\scri$ (infinity) with a $\emph{non-vanishing}$ $\Lambda\neq 0$. For $\Lambda>0$, some of the efforts on this topic can be found  in \cite{Ashtekar2014,Chrusciel2016,Ashtekar2017,Szabados2013,Fernandez-Alvarez-Senovilla2020b,Fernandez-Alvarez-Senovilla2022b} and references therein. The question for $\Lambda<0$ has also been  explored using different approaches, both from  the relativistic \cite{Compere2019,Poole2019,KrtousPodolsky2004} and fluid/gravity  correspondence perspectives \cite{BernardideFreitas2014,GathEtal2015,deHaro2008,Mukhopadhyay2013,Ciambelli2018}. Nevertheless, a general satisfactory result determining the presence of gravitational radiation was missing so far.\\

In a series of works \cite{Fernandez-Alvarez-Senovilla2020a,Fernandez-Alvarez-Senovilla2020b,Fernandez-Alvarez-Senovilla2022a,Fernandez-Alvarez-Senovilla2022b,Fernandez-Alvarez-Senovilla2025}, we have established a criterion that identifies the existence of gravitational radiation at $\scri$ for arbitrary $\Lambda$. Our method is based on superenergy (or tidal energy) methods ---see \cite{Senovilla2000} and references included there---  and conformal completions à la Penrose \cite{Penrose63}. The idea is that tidal energy flux is identified with the existence of gravitational radiation, and this is measured by the Weyl curvature. For a given observer one defines the super-Poynting vector ---introduced in \cite{Bel1958,Bel1962} and studied for its relevance on characterising intrinsic gravitational radiation  \cite{Bel1962,Pirani57,Wheeler77,Zakharov}---, in analogy with the electromagnetic Poynting vector ---see also \cite{Senovilla2000,Alfonso2008}. Based on the existence of a flux of tidal energy at infinity, our criterion identifies the presence of gravitational radiation there \cite{Fernandez-Alvarez-Senovilla2020b,Fernandez-Alvarez-Senovilla2022b,Senovilla2022} for any $\Lambda$. This has been shown to be equivalent to the traditional news tensor in $\Lambda=0$ space-times \cite{Fernandez-Alvarez-Senovilla2020a,Fernandez-Alvarez-Senovilla2022a} and application to a very general family of black hole solutions with $\Lambda >0$ have shown that gravitational radiation arrives at infinity if and only if the black holes are accelerated \cite{Fernandez-Alvarez-Podolsky-Senovilla2024} ---see also J. Podolsky's contribution to this session. In the $\Lambda<0$ scenario, our criterion fits neatly into Friedrich's initial boundary problem formulation  \cite{Friedrich1995} and gives rise to incoming/outgoing gravitational radiation conditions. Importantly, the results are fully covariant and gauge-independent.\\

 Our characterisation is formulated in the unphysical spacetime $ M $ with metric $\ct{g}{_{\alpha\beta}}$ related to the physical spacetime $(\hat{M} , \cth{g}{_{\alpha\beta}})$ by a conformal transformation  $\ct{g}{_{\alpha\beta}}=\Omega^2\cth{g}{_{\alpha\beta}}$. There is the gauge freedom $\Omega \rightarrow \omega \Omega$ with $\omega$ a positive function on $M$, and  \emph{all our results are gauge independent}. Infinity $\scri$ is represented by the boundary of $\hat{M}$ in $M$, located at $\Omega =0$. This is a hypersurface with lightlike (if $\Lambda$=0), spacelike (if $\Lambda>0$) or timelike (if $\Lambda<0$) character. The normal to $\scri$ is defined there as
 	\begin{equation}
 		\ct{N}{_{\alpha}}\defeq \partial_{\alpha}\Omega\ ,
 	\end{equation}
Remarkably, the Weyl tensor $\ct{C}{_{\alpha\beta\gamma}^{\delta}}$ vanishes at {$\scri$ ---ergo no super-Poynting there---}, so if one wants to study the asymptotic tidal distortion of the space-time caused by gravitational waves, one has to use the \emph{rescaled Weyl tensor}
 	\begin{equation}
 		\ct{d}{_{\alpha\beta\gamma}^{\delta}}\defeq\frac{1}{\Omega}\ct{C}{_{\alpha\beta\gamma}^{\delta}}\ .
 	\end{equation}
This tensor field is regular and non-vanishing in general  at $\scri$ ---see for instance \cite{Fernandez-Alvarez-Senovilla2022b,Kroon} and references therein for more details on this.

\section{The asymptotic super-momenta and super-Poynting}
	The central object in our analysis is the superenergy tensor of the rescaled Weyl tensor, the \emph{rescaled Bel-Robinson tensor}:
	\begin{equation}
		\D_{\alpha\beta\gamma\delta} :=  d_{\alpha\mu\gamma}{}^\nu d_{\delta\nu\beta}{}^\mu +d_{\alpha\mu\delta}{}^\nu  d_{\gamma\nu\beta}{}^\mu -\frac{1}{8} g_{\alpha\beta} g_{\gamma\delta} d_{\rho\sigma\mu\nu} d^{\rho\sigma\mu\nu}\ .
	\end{equation}
Given a casual vector $\ct{u}{^\alpha}$ at $\scri$, the \emph{asymptotic supermomentum} relative to $\ct{u}{^{\alpha}}$ is defined as
	\begin{equation}\label{s-mom}
		\Pi^\alpha (\vec u) := -u^\mu u^\nu u^\rho \ct{\D}{^{\alpha}_{\mu\nu\rho}} |_\scri \, .
	\end{equation}
 Furthermore, when $u^\alpha$ is timelike it can be chosen to be unit ---as we will always do--- and decomposing $\ct{\Pi}{^{\alpha}}\prn{\vec{u}}$ with respect to $\ct{u}{^{\alpha}}$, gives
	\begin{equation}\label{eq:supermomentum}
			\ct{\Pi}{^\alpha}(\vec u) = {\cal W}  \ct{u}{^{\alpha}} + \ct{\overline{\P}}{^\alpha} (\vec u) , \hspace{1cm} u_\mu \ct{\overline{\P}}{^\mu} =0.
	\end{equation}
Here, $\W$ and $ \ct{\overline{\P}}{^\alpha}$ are the \emph{asymptotic superenergy density} and \emph{asymptotic super-Poynting vector}, respectively.
\section{Characterization of radiation at $\scri$}
	\subsection{The general criterion}
	\Cref{eq:supermomentum} encodes the flux of tidal energy measured by an observer $\ct{u}{^{\alpha}}$. The question, then, is to identify the preferred observers at $\scri$ that detect the flux of superenergy traversing $\scri$. This task is summarised in the following general formulation:
		\begin{crit*}[\textbf{general formulation}]
			There is no gravitational radiation crossing an open portion $\Delta\subset \scri$ 
			if and only if the {\em flux of tidal energy traversing $\Delta$} ---measured by observers geometrically selected by the structure of $\scri$— vanishes.
		\end{crit*}

	This statement might look too abstract and observer dependent. However, it turns out that the criterion translates into a \emph{covariant condition,  independent of the choice of coordinates and gauge, given in terms of geometric objects at $\scri$}. Such relation takes a different form depending on the value of the cosmological constant, as we present next.
	\subsection{The $\Lambda=0$ case}
		This is the most studied case in the literature, also the one in which gravitational-wave phenomenology is typically considered. 
		The vanishing of $\Lambda$ makes $\scri$ a null hypersurface, and this provides us with \emph{the unique selected observer}: the normal to $\scri$, $\ct{N}{^\alpha}$. Observe that this vector field is lightlike, and as such is tangential to $\scri$. The asymptotic super-momentum for $\Lambda=0$ is then
			\begin{equation}
				\ct{\Pi}{^\alpha}\prn{\vec{N}}=-\ct{N}{^\mu}\ct{N}{^{\nu}}\ct{N}{^{\rho}}\ct{\D}{^{\alpha}_{\mu\nu\rho}}\ ,\quad\quad\ \ct{N}{^{\alpha}}\ct{N}{_{\alpha}}=0\ .
			\end{equation}
		 This supermomentum has the particular property of being a lightlike vector field itself, and its properties were first studied in  \cite{Fernandez-Alvarez-Senovilla2020a,Fernandez-Alvarez-Senovilla2022a} were we called it asymptotic \emph{radiant}\footnote{Here `radiant' means that the asymptotic supermomentum is built with a null vector field.} super-momentum. Its vanishing  gives the correct characterisation, providing an alternative to the traditional `news criterion':  ``no radiation if and only if News vanishes''.
		 We showed the following equivalence between both criteria:
		 	\begin{thm}[No radiation on $ \Delta\subset\scri$, case $\Lambda =0$] 
		 	There is no gravitational radiation on the open portion $ \Delta \subset \scri $ if and only if the asymptotic radiant super-momentum $ \Pi{^\alpha} (\vec N) $ vanishes on $\Delta$:
		 	\begin{equation}
		 	\mbox{ News vanishes on $\Delta$}  \quad\Longleftrightarrow \quad \Pi{^\alpha}(\vec N) \stackrel{\Delta}{=}  0 .
		 	\end{equation}
		 	\end{thm}	
Our new criterion serves to study the Petrov types of the re-scaled Weyl tensor directly as well as the relationship between the presence of radiation and the peeling-expansion terms of the Weyl tensor \cite{Fernandez-Alvarez-Senovilla2022c}.
	 As a bonus, it is easy to check if a given space-time  contains gravitational radiation at $\scri$  as one only needs to compute the asymptotic supermomentum: the rescaled Weyl tensor can be obtained with standard existing methods of computer algebra systems. This avoids the use of coordinate expansions or solving differential equations to compute the news tensor. Also, the asymptotic supermomentum was related to the  Bondi-Trautman energy loss \cite{Fernandez-Alvarez-Senovilla2020a,Fernandez-Alvarez-Senovilla2022a}, showing that an extra factor arises that puts it at the \emph{tidal-energy level} ($M T^{-2}L^{-3}$), and not at the energy-density level ($M T^{-2}L^{-1}$), as it has to be.
	
	\subsection{The $\Lambda>0$ case}\label{ssec:lambda-positive}
	This is the case with a spacelike $\scri$ and, thus, with a timelike normal $\ct{N}{^{\alpha}}$  that we normalize
	\begin{equation}
			\ct{n}{_{\alpha}}\defeq\sqrt{\frac{3}{\vert\Lambda\vert}}\ct{N}{_{\alpha}}\ .
		\end{equation}
		This is the \emph{unique geometrically selected} causal observer at $\scri$, and the asymptotic supermomentum 
		 is given now by \eqref{s-mom} with $u^\alpha \rightarrow n^\alpha$. 
	The flux of asymptotic tidal energy crossing $\scri$ is then given by the \emph{asymptotic super-Poynting vector} field $\ct{\overline{\P}}{^\alpha} \prn{\vec n}$. Using $\ct{n}{_{\alpha}}$, the standard electric and magnetic parts of the rescaled Weyl tensor are defined, respectively, as
		\begin{align}
			\ct{D}{_{\alpha\beta}}&\defeq\ct{n}{^{\mu}}\ct{n}{^{\nu}}\ct{d}{_{\alpha\mu\beta\nu}}\ ,\label{eq:electric} \\
			\ct{C}{_{\alpha\beta}}&\defeq\ct{n}{^{\mu}}\ct{n}{^{\nu}}\ctr{^*}{d}{_{\alpha\mu\beta\nu}}\ . \label{eq:magnetic}
		\end{align}
	where $^*$ is the Hodge star operator. On the one hand, the tensor $\ct{D}{_{\alpha\beta}}$ is also the $3$-rd order coefficient (in $3+1$ dimensions) of the Fefferman-Graham expansion \cite{Fefferman-Graham-1985,Fefferman-Graham-2011}. On the other hand, $\ct{C}{_{\alpha\beta}}=-\sqrt{{3}/{\Lambda}}\ct{Y}{_{\alpha\beta}}$, where $\ct{Y}{_{\alpha\beta}}$ is the Cotton-York tensor of 
	$\scri$, hence $\ct{C}{_{\alpha\beta}}$ depends on the intrinsic conformal geometry of $\scri$ only. Furthermore, results by Friedrich \cite{Friedrich1986a,Friedrich1986b} established that
	a  three-dimensional Riemannian manifold $\scri$ endowed with a conformal class of metrics 
	and  a traceless divergence-free  tensor  $\ct{D}{_{\alpha\beta}}$ conforms initial/final data for the $\Lambda>0$-vacuum Einstein field equations. Thus, any information about gravitational radiation should take into account both $\ct{D}{_{\alpha\beta}}$ and $\ct{C}{_{\alpha\beta}}$. Indeed, the asymptotic super-Poynting can be written as\footnote{Here $\ct{\epsilon}{_{\alpha\beta\gamma}}$ is the three-dimensional volume form at $\scri$.}
		\begin{equation}
			 \ct{\overline{\P}}{_{\mu}}\prn{\vec{n}} := 2\epsilon_{\mu\nu\rho} C^{\nu\sigma} D^\rho{}_\sigma\ ,
		\end{equation}
	hence our general criterion in this case translates to the following covariant statement
	\begin{thm}[No radiation on $ \Delta\subset\scri$, case $\Lambda >0$]
		There is no gravitational radiation on the open portion $ \Delta \subset \scri $ if and only if $\ct{D}{_{\alpha\beta}}$ and $\ct{C}{_{\alpha\beta}}$ commute.
	\end{thm}
	\subsection{The $\Lambda<0$ case}
	Finally, $\Lambda <0$ implies that $\scri$ is a timelike, Lorentzian,  3-dimensional manifold with spacelike unit normal $\ct{n}{_{\alpha}}$. The distinguishing feature of this case is that the geometry does \emph{not} determine a \emph{unique} causal observer. Instead, it selects a whole family of asymptotic observers $\ct{u}{^{\alpha}}$ that are \emph{tangent} to $\scri$ (i.e., $\ct{u}{^{\mu}}\ct{N}{_{\mu}}=0$), each one having its own supermomentum 
		\begin{equation}
				\Pi^\alpha (\vec u) := -u^\mu u^\nu u^\rho \ct{\D}{^{\alpha}_{\mu\nu\rho}}, \ \ \  u^{\mu}u_{\mu}=-1\ ,u^{\mu}N_{\mu}=0\ .
		\end{equation}
	 Then, there is no gravitational radiation \emph{traversing} $\scri$ if and only if \emph{all} asymptotic super-Poynting vectors relative to observers $\ct{u}{^{\alpha}}$ within $\scri$ do not have transversal component. That is
		\begin{equation}\label{eq:norad-AdS}
			 			N_\mu \ct{\overline{\P}}{^\mu} \prn{\vec{u}}=0\ \ \ \ \ \ \forall\ u^{\mu}, \ \ \  u^{\mu}u_{\mu}=-1\ ,u^{\mu}N_{\mu}=0\ .
		\end{equation}
	 The tensors defined by \cref{eq:electric,eq:magnetic} have now the same geometric meaning ($\ct{C}{_{\alpha\beta}}$ is the Cotton-York tensor of $\scri$, and $\ct{D}{_{\alpha\beta}}$ is part of the data in the Fefferman-Graham expansion) but are computed with respect to the \emph{spacelike} $\ct{n}{_{\alpha}}$ ---hence they are not the `magnetic' and `electric' parts of the rescaled Weyl tensor. Interestingly,  \eqref{eq:norad-AdS} translates into a covariant statement too \cite{Fernandez-Alvarez-Senovilla2025}:
	 \begin{thm}[No radiation on $\Delta\subset \scri$, case $\Lambda<0$]
	 	 		There is no gravitational radiation on an open portion $\Delta\subset \scri $ if and only if  $C_{\alpha\beta}$ and $D_{\alpha\beta}$ are functionally linearly dependent there, i.e., there exist real functions $\beta$ and $\gamma$ on $\Delta\subset \scri$ such that
	 \begin{equation}
		 \beta D_{\alpha\beta}=\gamma C_{\alpha\beta}\ , \hspace{5mm} \gamma^2+\beta^2 \neq 0 .
	 \end{equation}
 	\end{thm}
	 	 	
	 \begin{remark}
			 A variation of the criterion gives a condition for no incoming (outgoing) gravitational radiation:
			 $$
			 	\text{No incoming (outgoing) gravitational radiation } \iff 	N_\mu \ct{\overline{\P}}{^\mu} \prn{\vec{u}}\leq 0 \ \ \ (\geq 0) \ \ \ \ \ \ \forall\ u^{\mu}, \ \ \  u^{\mu}u_{\mu}=-1\ ,u^{\mu}N_{\mu}=0\ .
			 $$
	 \end{remark}
\section{Conclusion}
Our program of characterising the existence of gravitational radiation for arbitrary values of $\Lambda$ is now complete. Absence of radiation at $\Delta\subset \scri$ is summarised as follows:

\begin{center}
\begin{tabular}{ |c|c|c|c| } 
\hline 
& & &  \\[-0.5ex]
{ $\Lambda$} & \textbf{Super-momentum} & \textbf{Super-Poynting} & \textbf{Covariant formulation}\\
& & &\\[-0.6ex]
\hline
& & &  \\[-0.5ex]
$ \Lambda = 0$ & $\Pi^\mu (\vec N)=0$  &  & $N^\mu$ is a multiple PND  of $\ct{d}{_{\alpha\beta\gamma}^{\delta}}$\\
& & &  \\[-0.5ex]
\hline
& & &  \\[-0.5ex]
$ \Lambda > 0 $ &  $\Pi^\mu(\vec n) = \W(\vec{n}) n^\mu$ &  ${\overline{\cal P}}^\mu(\vec n) =0$ & $C_{\mu\nu}(\vec n)$ and  $D_{\mu\nu}(\vec n)$ commute \\
& & &  \\[-0.5ex]
\hline
& & &  \\[-0.5ex]
$ \Lambda < 0$ & $N_\mu \Pi^\mu({ \vec u}) =0,\quad$ $\forall \vec{u} \ \  N_\mu u^\mu =0$ & $N_\mu \overline{\cal P}^\mu({\vec u}) =0\quad$ $\forall \vec{u} \ \  N_\mu u^\mu =0$ & $\beta D_{\mu\nu} = \gamma C_{\mu\nu}$ for $\beta$, $\gamma$ functions \\[1ex] 
 \hline
\end{tabular}
\end{center}
\section*{Acknowledgements}
FFÁ was partly supported by the Grant Margarita Salas MARSA22/20 (Spanish Ministry of Universities and European Union), financed by European Union --Next Generation EU-- and by a contributory benefit from the Spanish SEPE (Servicio Público de Empleo Estatal).  JMMS is supported by the Basque
Government grant number IT1628-22, and by Grant PID2021-123226NB-I00 funded by the Spanish
MCIN/AEI/10.13039/501100011033 together with ``ERDF A way of making Europe''.
\bibliographystyle{iopart-num}
\bibliography{gr24-fernandez-and-senovilla.bib}

\end{document}